\documentclass{evn2004}
\usepackage{txfonts}
\usepackage{graphicx}
\begin{document}
\title{Tracking the curved jet in PKS 1502+106}
%   \title{Extreme Superluminal Motion in the Curved Jet of PKS\, 1502+106}

\author{ T.~An      \inst{1,2}
        \and X.Y. Hong  \inst{1,2}
        \and T. Venturi \inst{3}
        \and D.R. Jiang \inst{1,2}
        \and W.H. Wang  \inst{1,2} }%

\institute{Shanghai Astronomical Observatory, Chinese Academy of Sciences, Shanghai
200030, China \and National Astronomical Observatories, Chinese Academy of Sciences,
Beijing 100012, China \and Istituto di Radioastronomia del CNR, Via Gobetti 101, 40129
Bologna, Italy }

   \abstract{
We carried out a multifrequency and multiepoch study of the highly polarized quasar,
PKS\, 1502+106 at radio frequencies. The analysis is based on an EVN dataset at 5 GHz,
archive VLBA datasets at 2.3, 8.3, 24.4 and 43.1 GHz and an archive MERLIN dataset at 5
GHz. The various datasets span over a period of 10 years.
The source is characterized by a multi--component
one--sided jet at all epochs. The VLBI images show that a complex curved jet is located
to the southeast and east of the core, with the position angle (PA) of the jet axis
wiggling between $\sim$80\degr{} and $\sim$130\degr{}. The MERLIN image reveals that the
jet extends to 0.6\arcsec at a PA $135\pm12$\degr{}. The radio core in the VLBI images has
a brightness temperature approaching the equipartition limit, indicating highly
relativistic plasma beamed towards us. $\Delta PA$ in the source, the misalignment of the
kpc-- and pc--scale radio structure, is estimated about 32\degr{}, suggesting that PKS
1502+106 belongs to the aligned population. Four superluminal components are detected in
the parsec scale jet, whose velocities are 24.2$h^{-1}c$, 14.3$h^{-1}c$, 6.8$h^{-1}c$ and
18.1$h^{-1}c$. Our analysis supports the idea that the relativistic jet in
\object{PKS\,1502+106} is characterised by extreme beaming and that its radio properties
are similar to those of $\gamma$--ray loud sources. }

\maketitle
%
%________________________________________________________________

\section{Introduction}

One of the most significant observational results of extragalactic $\gamma$--ray active
galactic nuclei (AGNs) is that all EGRET--identified objects are radio--loud sources
(Mattox et al. \cite{Mattox}). Relativistic beaming in the jet is used to explain the
EGRET identification in radio--loud AGNs. The EGRET--identified sources have on average
much faster apparent superluminal motions than the general population of radio--loud
sources (Jorstad et al. \cite{Jorstad}). Hong et al. (\cite{Hong}) concluded that
$\gamma$--ray loud quasars typically show aligned morphologies on parsec and kiloparsec
scales based on a statistical analysis of the $\Delta PA$ distribution.
However, it is still a
matter of debate if the $\gamma$--ray emission in AGNs is related to higher beaming in
these sources.

PKS\,1502+106  (4C 10.39, OR103), $z$=$1.833$ (Veron-Cetty \& Veron \cite{Veron}), is a
18.6 mag highly polarized quasar (Hewitt \& Burbidge \cite{Hewitt}; Tabara \& Inoue
\cite{Tabara}). A high and variable degree of polarization in the optical band is
reported by Impey \& Tapia (\cite{Impey}). It is known to be active and variable at
radio, optical and X--ray wavelengths (George et al. \cite{George} and references
therein). A VLA image at 1.64 GHz (Murphy, Browne \& Perley \cite{Murphy}) shows that a
continuous jet extends to the southeast and leads to a lobe located $\sim$~7\arcsec from
the core. VLBI observations (Fey, Clegg \& Fomalont \cite{Fey}; Fomalont et al.
\cite{Fomalont}; Zensus et al. \cite{Zensus}) exhibit a well--defined jet starting to the
southeast and sharply bending to the east at a distance of 3 -- 4 mas from the core.

Our interest in PKS\,1502+106 is related to the misalignment between the pc-- and
kpc--scale radio structure in AGNs and its relation to the $\gamma$--ray emission. A flux
density upper limit of $7 \times 10^{-8}$ photons cm$^{-2}$s$^{-1}$ (2$\sigma$) at
$\gamma$--ray energies was given by EGRET in the Phase I observation (Fichtel et al.
\cite{Fichtel}), while no detection is reported in in the following EGRET observations
(Thompson et al. \cite{Thompson}; Hartman et al. \cite{Hartman}). The gamma--ray
properties of this source are therefore unclear. We observed PKS 1502+106 with the EVN as
a gamma--ray source candidate, as part of a project (Hong et al. \cite{Hong02}) aimed at
studying the relation between the $\Delta PA$ and $\gamma$--ray emission . More details
are given in An et al. (\cite{An}).

%__________________________________________________________________
%__________________________________________________ One column table
   \begin{table*}
      \caption[]{Parameters of the images}
         \label{ImaPar}
\begin{tabular}{ccccccccc}\hline\hline
Figure & Epoch  &Band & Freq. & Beam              & S$_P$ &\emph{r.m.s.} & Contour &Ref.\\
       &        &     &(GHz)  &(mas)              & (Jy/b)&(mJy/b) &(mJy/b)\\\hline
Fig.1a & 1985.10&L    & 1.64  & 1510$\times$1480,48.6\degr & 2.20  &0.1    &0.44$\times$(-1,1,2,...,512)&1\\
Fig.1b & 1992.37&C    & 5.0   &79$\times$49,24.4\degr      & 1.63  &0.1    &3.0$\times$(-1,1,2,...,1024)&2   \\
Fig.1c & 1994.52&X    & 8.3   &7.18$\times$3.78,-1.7\degr  & 1.81  &1.1     &3.5$\times$(-1,1,2,...,128)&3\\
Fig.1d & 1997.85&C    & 5.0   &1.36$\times$1.21,64.7\degr  & 0.80  &0.7     &3.1$\times$(-1,1,2,...,256)&3\\
Fig.1e & 2002.65&K    &22.4   &0.64$\times$0.28,0\degr     & 0.73  &0.9     &2.8$\times$(-1,1,2,...,256)&3 \\
Fig.1f & 2002.37&Q    &43.1   &0.37$\times$0.16,0\degr     & 0.82  &1.0     &3.0$\times$(-1,1,2,...,256)&3\\
\hline\end{tabular}\\[2mm]
Ref: 1-- Murphy, Browne \& Perley \cite{Murphy}; 2-- archival MERLIN data; 3-- An, et al.
\cite{An}
\end{table*}

   \section{Results}

Figure 1 shows images of PKS 1502+106 made with the VLA, MERLIN, EVN and VLBA,
illustrating the complex curved jet morphology on very different scales. Figure 2
presents the distribution of jet position angles along the projected radial distance in
PKS 1502+106:

\begin{itemize}

    \item The VLA map in Fig. 1a (Murphy, Browne \& Perley \cite{Murphy}) shows a strong core and
    a straight jet directed to the southeast. The core is unresolved at the resolution
    of 3\arcsec.
    A low-brightness lobe is located 7\arcsec{} from the core, in a PA $\sim$157\degr{}. No
    counterjet is found at the dynamic range of of the image, i.e. 5000:1.

    \item The MERLIN image in Fig. 1b, obtained with the archive 5 GHz data, exhibits the
    jet extending to the southeast up to 0.6\arcsec{} from the core. The jet starts out
    from the core at PA$\sim$120\degr{} within 0.1\arcsec{}, and then gradually bends in
    the direction of the arcsec--scale lobe (7\arcsec{} from the core) after a curve
    at $\sim0.3\arcsec$. The jet is resolved into a number of knots.

    \item The VLBI images displayed in Figs. 1c to 1f, obtained with the EVN and VLBA,
    show a complex curved jet with a length of 15 mas. Several distinct components
    follow a continuous curved path. The innermost component within 0.2 mas shows no radial motion
    but significant changes in position angle, and the components from 0.5 mas to 4 mas
    move out to the southeast at a PA$\sim125\degr$ (Figure 2). There is a clear bending at
    3 -- 4 mas, from PA$\sim130\degr$ to PA$\sim80\degr$ (Figure 1 and Figure 2).

    \item Considering the complex parsec--scale jet structure, we define the jet axis PA
    on parsec scale as the PA measured at 10 pc ($\sim1.6$ mas at z=1.833)
    from the core, i.e. $\sim125\degr{}$.
    The jet PA on kpc scale is defined as the position angle of the lobe in the VLA image,
    $\sim$157\degr{}. Hence, $\Delta PA$ is 32\degr{}, indicating that PKS 1502+106 an aligned
    quasar.
\end{itemize}

   \begin{figure*}
   \centering
   \includegraphics[width=0.8\textwidth]{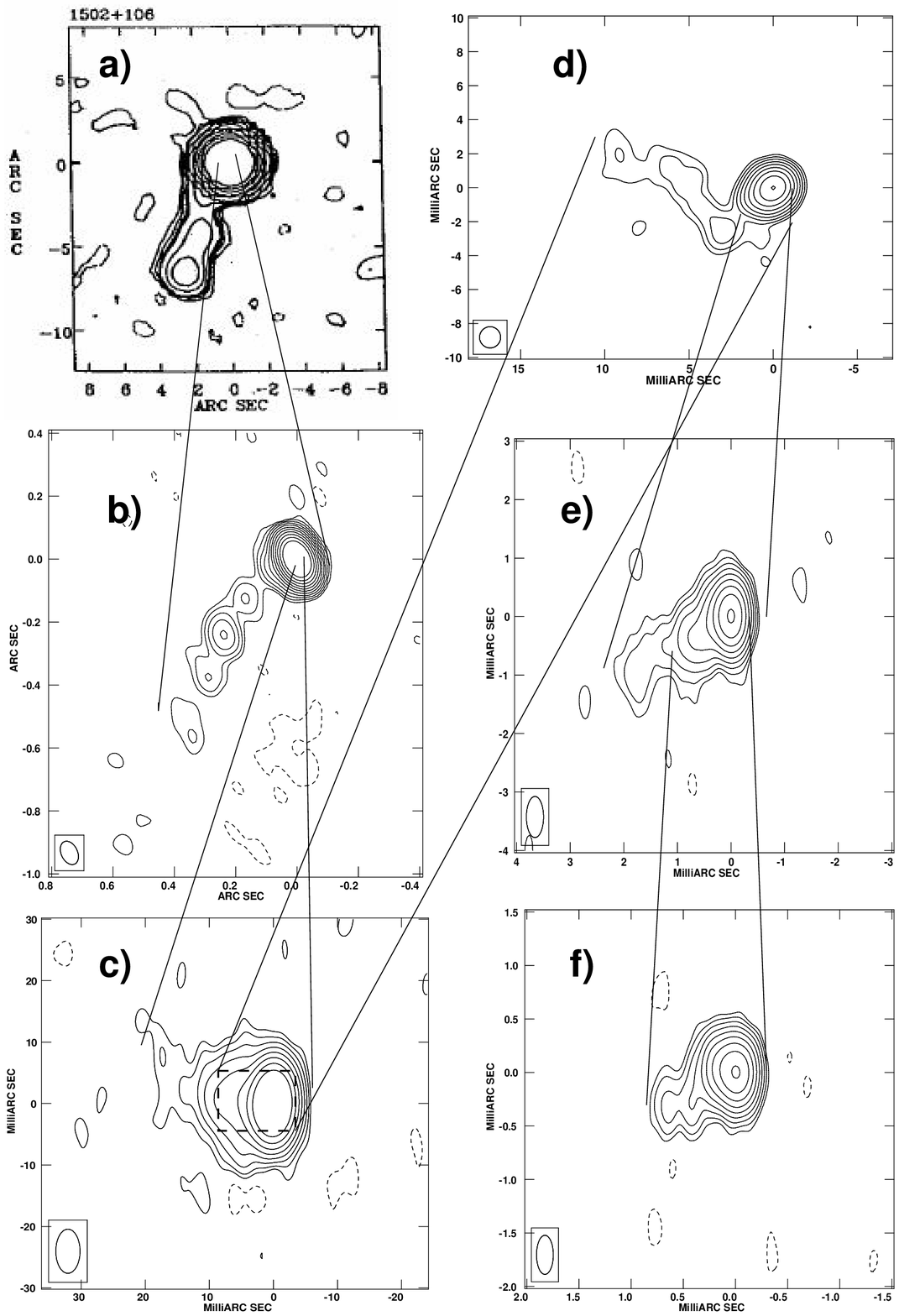}
      \caption{Total intensity images of PKS 1502+106.
      Image parameters are referred to Table \ref{ImaPar}.
      VLA image (Murphy, Browne \& Perley \cite{Murphy});
      MERLIN image (archive 5 GHz data); others (An et al. \cite{An}).}
         \label{CLEAN}
   \end{figure*}

  \begin{figure}
   \centering
   \includegraphics[width=0.5\textwidth]{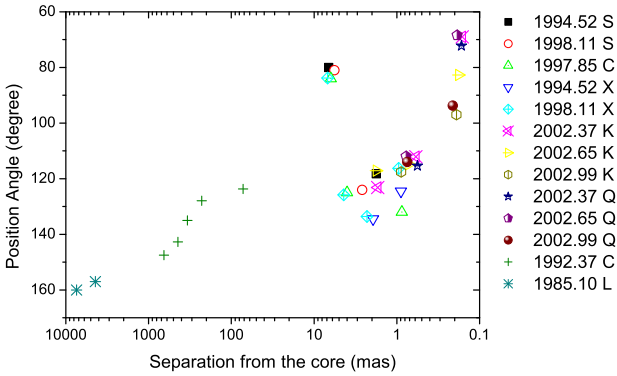}
      \caption{Position angles of jet components along the radius.}
         \label{PAR}
   \end{figure}

The compact core component in the VLA and MERLIN images contains more than 98\% of the total
flux density, indicating strong Doppler beaming effect.

The multifrequency VLBI datasets allow us to estimate the spectral index
($S_\nu\propto\nu^\alpha$) distribution of the source. As expected, the shape of the
spectrum differs in the core and in the jet components. The spectrum is flat in the core
($\alpha_{43}^{24} \sim 0$), while it is in the range $-0.7 \div -1.5$ along the inner
jet. We note that the spectrum at epoch 2002.99 is steeper in each component compared
to other epochs. It could be related to the jet expansion. We also estimated the
brightness temperature of the core based on 5, 8.3, 24.4 and 43.1 GHz VLBI observations.
The result gives $T_r=(2.0\pm0.5)\times 10^{11}$K in the source frame. It approaches the
upper limit constrained by equipartition (Kellermann \& Pauliny-Toth \cite{Kellermann};
Readhead \cite{Readhead}), suggesting highly relativistic plasma close to the
line--of--sight.

We detected four  superluminal components along the parsec--scale jet, whose velocities
are 24.2$h^{-1}c$, 14.3$h^{-1}c$, 6.8$h^{-1}c$ and 18.1$h^{-1}c$, respectively. The
speeds are much higher than the average level of radio loud quasars (3.2c, Pearson et al.
\cite{Pearson}), and are found in agreement with the average speed of gamma--ray loud
quasars (15.9$\pm$6.6$h^{-1}c$, Jorstad et al. \cite{Jorstad}). We note that the highest
superluminal motion is associated with a component passing through the jet bending at
3--4 mas from the core. The apparent motion could be magnified to this even higher level
by projection effects.

\section{Conclusions and Future Work}

We carried out a mutlifrequency and multiepoch analysis of PKS 1502+106 at mas and arcsec
resolutions. The source morphology is highly variable and the jet structure is very
complex. The jet PA wiggles in 80\degr{}--130\degr{}--80\degr{} in the inner 20 mas, and
changes to 120\degr{}--160\degr{} on a scale of hunderds of mas to several arcsec. The
overall oscillated jet trajectory could be described by a helical pattern. We note that
the wide range of PA on the mas scale could be the result of amplification by projection
effects due to a jet alignment close to the line--of--sight. Similar wiggling structures
have also been found in other extragalactic jets, such as 3C 273, 3C 345, 3C 120, MrK
501, M87. They could be interpreted as motions along a helix lying on the surface of a
cone. Helical trajectory in relativistic jets could originate from regular precession of
the jet flow at the central engine, or from random disturbance at the jet base. The
initial disturbed jet might evolve into helical motion at larger scales (Hardee
\cite{Hardee} \cite{Hardee03}). Other dynamical processes, such as bent shocks, could also
result in a projected helical path (G\'{o}mez, Alberdi \& Marcaide \cite{Gomez}). PKS
1502+106 could be a good case for a morphological study of helical jets.

Most of the total flux density in the VLA and MERLIN images is dominated by an unresolved
core component; the brightness temperature of the radio core on the parsec scale
approaches the equipartition limit; four superluminal components with extremely high
apparent speed are detected in the parsec--scale jet. All the observational results
support the hypothesis of a highly relativistic jet flow aligned towards
us. \\
The properties of $\Delta PA$ ($\sim$30\degr{}) and superluminal motion in PKS
1502+106 are consistent with those of $\gamma$-ray loud quasars (Hong et al. \cite{Hong};
Jorstad et al. \cite{Jorstad}). These observed features, together with the uncertain
$\gamma$--ray emission in this source, pose again the question of the intrinsic difference
between $\gamma$--ray loud and quiet extragalactic radio sources. A confirmation of
$\gamma$-ray emission from this source would be highly valuable for our understanding of
the $\gamma$-ray loudness phenomenon in radio loud quasars.

%  \begin{figure*}
%   \centering
%   \includegraphics[width=1.0\textwidth]{AnT-3.eps}
%      \caption{2-D projected jet structure of PKS 1502+106.
%      The signs are referred to those in Figure 2. The "cross" marks
%      the position of the core.}
%         \label{RADEC}
%   \end{figure*}

\begin{acknowledgements}
This research was supported by the National Science Foundation of PR China (10333020,
10328306 and 10373019). The authors thank A.L. Fey for providing archival VLBA datasets,
and Anita M. S. Richards for providing archival MERLIN dataset. T. An acknowledges the support
of the European Commission's I3 Programme ``RADIONET", under contract No.\ 505818 to attend the conference.
The European VLBI Network is a joint facility of
European, Chinese, South African and other radio astronomy institutes funded by their
national research councils.
\end{acknowledgements}

%______________________________________________________________


\begin{thebibliography}{}
\bibitem[2004]{An} An, T., Hong, X.Y., Venturi, T., et al. 2004, A\&A, 421, 839
\bibitem[1996]{Fey} Fey, A. L., Clegg, A. W., \& Fomalont, E. B. 1996, ApJS, 105, 299
\bibitem[1994]{Fichtel} Fichtel, C.E., Bertsch, D.L., Chiang, J., et al. 1994, ApJS, 94, 551
\bibitem[2000]{Fomalont} Fomalont, E.B., Frey, S., Paragi, Z., et al. 2000, ApJS, 131, 95
\bibitem[1994]{George} George, I.M., Nandra, K., Turner, T.J., et al. 1994, ApJ, 436L, 59
\bibitem[1994]{Gomez}G\'{o}mez, J.L., Alberdi, A. \& Marcaide, J.M. 1994, A\&A, 284, 51
\bibitem[1987]{Hardee} Hardee, P.E. 1987, ApJ, 318, 78
\bibitem[2003]{Hardee03} Hardee, P.E. 2003, ApJ, 597, 798
\bibitem[1999]{Hartman} Hartman, R.C., Bertsch, D.L., Bloom, S.D., et al. 1999, ApJS, 123, 79
\bibitem[1989]{Hewitt} Hewitt, A., \& Burbidge, G. 1989, A New Optical Catalog of
Quasi-Stellar Objects (Chicago: Univ. Chicago Press)
\bibitem[1998]{Hong} Hong, X.Y., Jiang, D.R., \& Shen, Z.Q. 1998, A\&A, 330, L45
\bibitem[2002]{Hong02} Hong, X.Y., et al. 2002, Proceedings of 6th European VLBI Network
Symposium, eds. by E.Ros, R.W.Porcas, A.P.Lobanov, and J.A.Zensus, p.103
\bibitem[1988]{Impey}Impey, C.D., \& Tapia S. 1988, ApJ, 333, 666
\bibitem[2001]{Jorstad}Jorstad, S.G., Marscher, A.P., \& Mattox, J.R., et al. 2001, ApJS, 134, 181
\bibitem[1969]{Kellermann}Kellermann, K.I., \& Pauliny-Toth, I.I.K. 1969, ApJ, 155, L71
\bibitem[1997]{Mattox} Mattox, J.R., Schachter, J., Molnar, L., et al. 1997, ApJ, 481, 95
\bibitem[1993]{Murphy} Murphy, D.W., Browne, I.W.A., \& Perley, R.A.
    1993, MNRAS, 264, 298
\bibitem[1998]{Pearson} Pearson, T.J., Browne, I.W.A., Henstock, D.R., et al. 1998, in ASP Conf. Ser.
    144, Radio Emission from Galactic and Extragalactic Compact Sources, ed. J.
    A. Zensus, G. B. Taylor, \& J. M. Wrobel (San Francisco: ASP), 17
\bibitem[1994]{Readhead} Readhead, A.C.S. 1994, ApJ, 426, 51
\bibitem[1980]{Tabara} Tabara, H., \& Inoue, M. 1980, A\&AS, 39, 379
\bibitem[1995]{Thompson} Thompson, D.J., Bertsch, D.L., Dingus, B.L., et al. 1995, ApJS, 101, 259
\bibitem[1998]{Veron} Veron-Cetty, M.P., \& Veron, P. 1998, ESO Sci. Rep., 18, 1
\bibitem[2002]{Zensus}Zensus, J.A., Ros, E., Kellermann, K.I., et al. 2002, AJ, 124, 662
\end{thebibliography}
\end{document}